# Spin-orbit interaction in InAs/GaSb heterostructures quantified by weak antilocalization

F. Herling[1,2], C. Morrison[1,3], C.S. Knox[1,4], S. Zhang[5], O. Newell[3], M. Myronov[3], E.H. Linfield[4] and C.H. Marrows[1]

[1]School of Physics and Astronomy, University of Leeds, Leeds, LS2 9JT, UK

[2]Novel Materials Group, Institut für Physik, Humboldt-Universität zu Berlin, 12489 Berlin, Germany

[3]Department of Physics, University of Warwick, Coventry, CV4 7AL, UK

[4]School of Electronic and Electrical Engineering, University of Leeds, Leeds, LS2 9JT, UK

[5]EPSRC National Centre for III-V Technologies, Department of Electronic and Electrical Engineering, University of Sheffield, Sheffield, UK

Abstract: We study the spin-orbit interaction (SOI) in InAs/ GaSb and InAs quantum wells. We show through temperature- and gate-dependent magnetotransport measurements of weak antilocalization that the dominant spin-orbit relaxation mechanism in our low-mobility heterostructures is Elliott-Yafet and not Dyakonov-Perel in the form of the Rashba or Dresselhaus SOI as previously suggested. We compare our findings with recent work on this material system and show that the SOI length lies within the same range. The SOI length may be controlled using an electrostatic gate, opening up prospects for developing spintronic applications.

InAs/GaSb heterostructures have seen renewed research interest in recent years. Early work focussed on coupling a 2D electron gas (2DEG) in the InAs layer with a 2D hole gas in the GaSb layer, creating hybridized electron-hole quantum states [1-8]. Much later, it was predicted that a topologically insulating state could be created in an InAs/GaSb heterostructure [9] by using band inversion due to the band edge positions of the conduction and valence bands in InAs and GaSb, respectively, to create an energy gap in the bulk and counter-propagating spin-polarized helical edge states, known as the quantum spin Hall effect (QSHE). Shortly after this theoretical prediction the bulk topological gap was demonstrated experimentally [10], and recent research has focussed on demonstrating spin polarized edge transport and confirmation of the existence of the QSHE [11-20].

The Rashba spin-orbit interaction (SOI) has been studied in 2DEGs in several InAs and InGaAs quantum well (QW) heterostructures [21-28]. The SOI gives rise to the conventional spin Hall effect (SHE), which is useful as a means of manipulating spin in a spintronic device, whilst its inverse can be used for detecting spin-polarized currents such as those in the helical edge states generated by the QSHE. The use of the SHE to detect such helical edge states in this manner was recently achieved in an HgTe based topological insulator [29], and a similar experimental approach could be employed in the InAs/GaSb system. One of the underlying causes of spin relaxation is the Dyakonov-Perel mechanism, which originates in the spatial inversion asymmetry of the heterostructure [30]. Another contribution is momentum scattering from phonons and impurities, known as the Elliot-Yafet (EY) mechanism, owing to the mixing of spin-up and spin-down states by the SOI of the lattice ions [31]. It is important to understand the effect of the proximity of the QWs to each other, and any induced interfacial effects, on the strength of the SOI and the relative contributions of momentum scattering (EY) and the SOIs (DP) arising from inversion asymmetry (bulk and structural).

Here we show, through magnetotransport weak antilocalization measurements of the temperature and gate dependence of the relevant scattering length scales that the dominant underlying spin relaxation mechanism in our InAs/GaSb heterostructure is EY, and not DP in the form of the Rashba SOI owing to structural inversion symmetry, as previously suggested [22-24,26]. We do this by demonstrating that the length for spin relaxation may be tuned using an electrostatic gate controlling the mean free path. Finally, we report our results obtained from an InAs QW, which also shows strong indicators for the EY mechanism.

We grew InAs/GaSb heterostructures using solid source molecular beam epitaxy on (100) GaAs substrates, with varying thicknesses of the InAs and GaSb layer, and varying AlSb barriers. A buffer of AlSb/GaAlSb was used to relax lattice mismatch strain and provide a pseudo-substrate for growth of the electrically active QW layers. The structures are capped with GaSb to prevent oxidation. The layer structure of the two heterostructures studied here are shown in figure 1; they are labelled A and B and are thus referred to throughout.

Magnetotransport measurements were performed in the temperature range 0.3 to 1.5 K using an Oxford Instruments Heliox AC-V $^3$He system with a 12T superconducting magnet, and at temperatures above 1.5 K, an Oxford Instruments $^4$He flow cryostat with an 8T superconducting magnet. In all measurements, the field was applied perpendicular to the plane of the device. Measurements were performed using a 50 µm wide and 250 µm long Hall bar geometry, fabricated by optical lithography, using a wet etching process to define the mesa, and 100 nm thick AuGeNi to form contacts that are Ohmic at all temperatures regardless of prior annealing. The top gate stack comprises a 30 nm thick layer of $Al_2O_3$ with a 100 nm thick Cr/Au electrode. The back gate was fabricated by depositing and annealing AuGeNi on the highly doped substrate. An optical microscope image of a typical device is shown in an inset to figure 2. In addition, square and Greek cross van der Pauw geometries were used. Electrical measurements were performed using an ac current excitation of 1 µA or 100 nA at 119.77 Hz and voltages were measured using Stanford Research Systems Model 830 lock-in amplifiers.

Carrier densities and mobilities at 1.7 K are given in Figure 1 for both wafers. The carrier density was calculated from the Hall voltage, which was fully linear for low fields and displays plateaus for high fields in accordance with the SdH oscillations. The majority carriers are electrons, determined from the sign of the Hall coefficient. There was no evidence of band inversion for top, back or double gated devices, perhaps due to the high carrier density and low mobility, so we cannot contribute to current arguments regarding recent findings for SOI close to the charge neutrality point [32].

Both wafer A and B exhibit weak antilocalisation (WAL). Figure 2 depicts the magnetoresistance for wafer A, in which there is a GaSb layer next to the InAs QW between the AlSb barriers. At low fields the behaviour shows the characteristic dip of WAL, which may be seen more clearly in the inset, with transitions to weak localisation (WL) and negative magnetoresistance above 500 mT, which suggests that a high density of impurities is present [30]. This would also explain the low mobility and high carrier density across all devices fabricated from this wafer. Other reasons for the unusually low mobility could be lattice mismatch or other interface effects.

Angular dependent measurements showed that the WAL and the Hall voltage are a sinusoidal function of the angle and vanish for an in-plane magnetic field, which demonstrates the two-dimensional nature of the carrier confinement within the QW.

The magnetoconductivity for all devices for different temperatures and applied gate voltages was fitted with the Hikami-Larkin-Nagaoka (HLN) model [33], which gives us three fitting parameters as characteristic lengths: the SOI length, $l_{SO}$, which gives the average distance travelled by an electron before a flip in spin occurs, the mean free path, $l_e$, between elastic scattering events, and the phase coherence length, $l_\phi$, between inelastic scattering events. An example fit is shown in Figure 3. The fit matches the experimental data points for the low field range and results are independent of the chosen interval, as long as it stays in the area in which WAL is present (below 400 mT (A) or 200 mT (B)). Beyond that it deviates owing to the negative high-field magnetoresistance.

An advantage of low mobility samples lies in the distinct WAL of the magnetoresistance, which can be more precisely fitted to obtain characteristic lengths. Accordingly, wafer A shows smaller errors for the characteristic lengths than wafer B, which has a higher mobility and lower carrier density (see figure 1). A disadvantage is the damping of the Shubnikov-de Haas (SdH) oscillations for high fields, so that only one oscillation (or two for wafer B) is visible and we cannot report high field SOI values through measurements of beating in the SdH oscillations, another common magnetotransport technique [22,24,26,34]. However, for certain materials the values for SOI extracted at low field from WAL are more accurate than the values deduced from high field beating in the SdH oscillations as the latter include the Zeeman effect [25] or intersubband scattering [35] which causes additional uncertainty in the fast Fourier transform method used to determine the spin-split carrier densities [36].

The change in resistance due to WAL, i.e. the quantum correction to the conduction at low temperatures for low fields, shows a logarithmic temperature dependence (figure 3 inset), which is the expected dependence for WAL, due to the power law behaviour of the lifetimes (spin/elastic/inelastic scattering), which can be converted to the lengths reported here [37].

The temperature dependence of the three characteristic lengths for wafer A is shown in figure 4. For low temperatures (up to about 5 K), a $T^{-1/2}$ behaviour is seen in phase coherence length, with a linear increase in mean free path and SOI length. For higher temperatures (above 5 K) the latter two have a very weak temperature dependence and the phase coherence length decreases linearly with increasing temperature. This suggests that electron-electron interactions are the dominant inelastic scattering mechanism for temperatures below ~5 K and that the electron-phonon interaction dominates at temperatures above that value [38]. The temperature independence of the elastic and SOI lengths suggests that these scattering rates are dominated by impurities, which corresponds to the analysis of the magnetoresistance data and suggests that the EY mechanism dominates in this heterostructure. The exponential drop in mean free path for the lowest temperatures (below about 1 K) is paralleled by the SOI length, as expected for our conclusion, but cannot be explained, as, for example, a freezing out of the impurities would show an opposite trend.

The inset in figure 6 shows the equivalent temperature dependence for of the SOI length for wafer B. It follows the same behaviour in the temperature range from 1.6 K to 12 K as wafer A for higher temperatures. The characteristic lengths for both wafers at 1.7 K are listed in figure 1. In accordance with our interpretation of the magnetoresistance and the temperature behaviour, a weaker SOI is seen for the lower carrier density wafer (B), which is the expected trend for the EY mechanism, but could also be explained by the DP mechanism, which originates in the missing bulk inversion symmetry [30]. The longer elastic and inelastic scattering lengths coincide with the high carrier mobility and can be explained with a lower impurity concentration, which would also lead to a shorter SOI length in the EY mechanism.

With the double (top and back) gated Hall bar samples for wafer A, the electrical transport properties (Hall sheet carrier density and mean free path) could be varied by around ten percent. The bottom inset in figure 5 shows that as the external electric field is made more negative the SOI length becomes shorter due to a change in mobility. We interpret this negative dependence as another reason to disregard Dresselhaus SOI as a dominant factor in our samples, because the SOI length would then be independent of changes in the local electric field. This is consistent with single InAs QWs, where the Dresselhaus term was responsible for less than 5 % of the SOI [39]. The Rashba SOI length (structural or interface inversion symmetry) is theoretically predicted to be also controllable by a gate voltage, but previous experiments for InAs QWs did not provide a definite experimental confirmation of the sign of the correlation. They either did not vary carrier density [22] or did so by illumination with a light emitting diode, rather than applying a gate voltage [23]. Experiments with gate voltages either reported no gate dependence [24] or a negative dependence for top gate voltages [26] as in our samples. Most recently Kim et al. found that the dependence is determined by the QW potential gradient in accordance with theoretical predictions [21,28,40]. So the negative dependence we report here could be caused by DP with Rashba SOI or EY.

The top inset in figure 5 shows the calculated band diagram of the InAs/GaSb heterostructure of wafer A. We performed the calculation with nextnano³, a 3D Poisson and Schrödinger solver [41]. There is only a small positive difference in energy across the InAs layer of approximately 30 meV. It is unlikely that such a small gradient would cause Rashba SOI of the order we report here, and furthermore it would lead to a negative dependence of the SOI length on a top gate voltage. It would also lead to constant mobility [42], but we see a linear mobility dependence on gate voltage, most likely caused by scattering from impurities. We therefore conclude that the SOI length is varied due to changes in the mean free path tuned by an external gate voltage.

In figure 5, a linear dependence of $l_{SO}$ on $l_e$ is seen for wafer A as the gate voltage is varied, just as we would expect for the EY mechanism, because spin relaxation is induced by scattering. Therefore, the DP mechanism (arising from Dresselhaus and Rashba SOI) seems to be negligible, as this would show a dependence of $l_{SO} \sim l_e^{-1}$, because spin precession is restarted by scattering [30,43].

To compare our reported SOI length value with the Rashba parameter $\alpha$ from other experiments with InAs QWs, we used:

$$\alpha = \frac{\hbar^2}{m^* l_{SO}} \tag{1}$$

to calculate the SOI length for different sources. We note that there is a variation in the reported accompanying effective mass $m^*$. The results are presented in table 1 and show that SOI length due to the EY mechanism that we measure here is in the range of the reported Rashba SOI lengths and cannot be neglected in a thorough investigation of spin relaxation mechanisms.

Figure 6 shows the $l_{SO}$-$l_e$ curve for wafer B as the gate voltage is varied. The linear trend we see for our InAs/GaSb is repeated, which points to the EY mechanism. Our calculation show an even smaller potential difference inside the QW of 10 meV, which of course could be affected by a high level of impurities. In contrast to wafer A we do not see linear dependence of mobility and mean free path on gate voltage. An explanation could be the larger variation of the transport parameters by a higher gate voltage owing to a more stable back gate. Therefore $l_e$ is varied over a larger range and could show dependence outside of the narrow picture of figure 5. We conclude that it is reasonable to say that the EY mechanism plays an important role in our InAs heterostructures but further research is necessary.

In conclusion, we report measurements of the spin-orbit interaction (SOI) in InAs/GaSb and InAs quantum wells using weak antilocalisation. We conclude that the dominant spin relaxation mechanism at least in the InAs/GaSb heterostructure reported here is the Elliott-Yafet mechanism, which also contributes in our InAs sample. We dismiss inversion asymmetry effects (Dyakonov-Perel) which have been used as explanation previously and report that the SOI length is controllable by a gate voltage, which makes its application in spintronic devices possible. We also add to the ongoing discussion about gate controllable Rashba SOI by showing that a thorough experimental procedure has yet to be reported. To account for all spin relaxation mechanism the dependence of the SOI parameter on the mean free path, the gate voltage and the intrinsic carrier density have to be considered.

ACKNOWLEDGMENT

We acknowledge helpful discussions with S.F. Fischer and O. Chiatti. We also thank M. Rosamond for advice in all clean room work, J. Batley for assistance with low temperature measurements and G. Burnell for software support.  This work was supported by the Engineering and Physical Sciences Research Council, the EPSRC platform grant 'Spintronics at Leeds' (EP/M000923/1), the EPSRC National Centre for III-V Technologies and the Erasmus+ program (F. H.).

| Author | Min. $l_{SO}$ (nm) | $m^*$ |
|---|---|---|
| Luo et al. [22] | 154 | 0.055 |
| Heida et al. [24] | 317 | 0.040 |
| Grundler et al. [26] | 47 | 0.036 |
| Schierholz et al. [27] | 167 | 0.026 |
| Kim et al. (2014) [28] | 221 | 0.050 |
| Kim et al. (2013) [21] | 237 | - |
| Park et al. [39] | 473 | - |
| **InAs/GaSb (this work)** | **150** | **-** |
| **InAs (this work)** | **380** | **-** |

*Table 1: Comparison of the reported minimal SOI lengths derived from the largest spin splitting parameter α reported, calculated with the accompanying effective mass $m^*$ or the value taken from literature (m\* = 0.023 electron masses).*

|  | (A) InAs/GaSb | (B) InAs |
|---|---|---|
| $n$ in $10^{11}$ cm$^{-2}$ | $(19 \pm 1)$ | $(7.4 \pm 0.1)$ |
| $\mu$ in cm$^2$/Vs | $(2{,}900 \pm 100)$ | $(10{,}300 \pm 100)$ |
| $l_{SO}$ in nm | $(150 \pm 8)$ | $(380 \pm 30)$ |
| $l_e$ in nm | $(82 \pm 6)$ | $(150 \pm 20)$ |
| $l_\phi$ in nm | $(280 \pm 10)$ | $(950 \pm 50)$ |

Figure 1: Comparison of the two wafers studied. The substrates are n-doped, the InAs/GaSb wafer A has an interface doping with Si atoms of $10^{11}$ cm$^{-2}$. The charge carrier densities, mobilities, and characteristic lengths are given at 1.7 K.

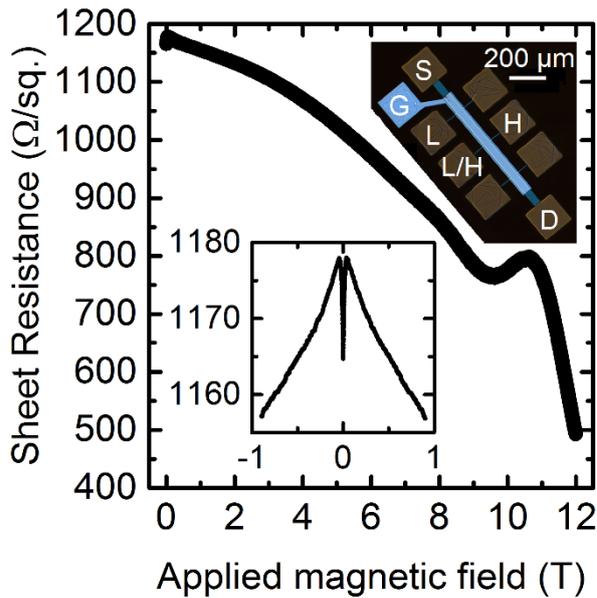

Figure 2: Magnetoresistance of an InAs/GaSb QW Hall bar sample (wafer A) at 340 mK and applied magnetic field of up to 12 T. Inset (top): Optical micrograph of the gated Hall bar geometry used here with labels for contacts (S = source, D = drain, G = gate, L = longitudinal voltage, H = Hall voltage). Inset (bottom): Low field magnetoresistance highlighting the WAL dip at zero field.

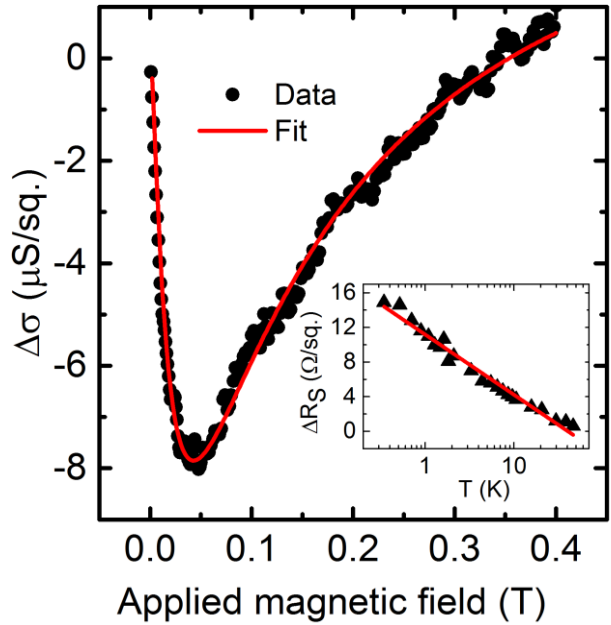

*Figure 3: The change in conductivity for an InAs/GaSb QW Hall bar sample (wafer A) at 1.7 K as a function of perpendicular magnetic field. The black data points show the measured values, the red solid line shows the HLN model fit with the following fit parameters: $l_{SO}$ = 147 nm, $l_e$ = 77 nm and $l_\varphi$ = 289 nm. The fit and data match for the pictured low field range, for higher fields the background negative magnetoresistance leads to deviation from this behaviour and is excluded from the fit. Inset: The absolute change in sheet resistance for low fields as a function of temperature on a log scale. The black data points show the measured values, the red solid line shows a linear fit.*

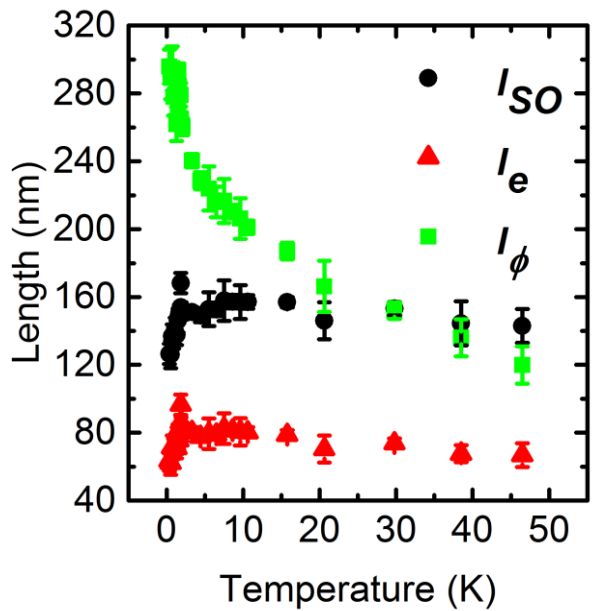

*Figure 4: Characteristic lengths from the HLN fit as a function of temperature for the InAs/GaSb wafer (A). In agreement with the model WAL only occurs for temperatures where $l_\varphi$ is of the order of $l_{SO}$ or bigger (around 50 K, see also the inset in figure 3).*

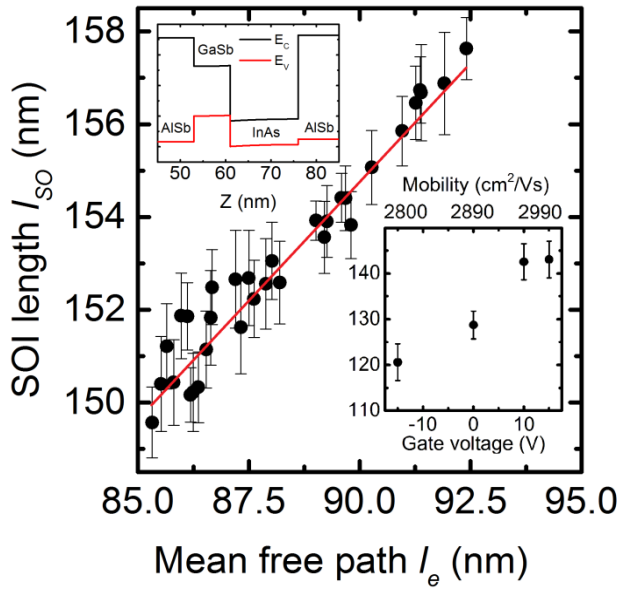

*Figure 5: The SOI length as a function of the mean free path controlled by an applied top and back gate voltage for wafer A at 1.7 K. Inset (top): Calculated band diagram showing the lowest conduction and the highest valence band energy, the band inversion inside the QW is also clearly evident. The Fermi level lies between the two band edges and only the first subband is populated. Inset (bottom): The SOI length as a function of the Hall mobility controlled by an applied top gate voltage for wafer A at 340 mK.*

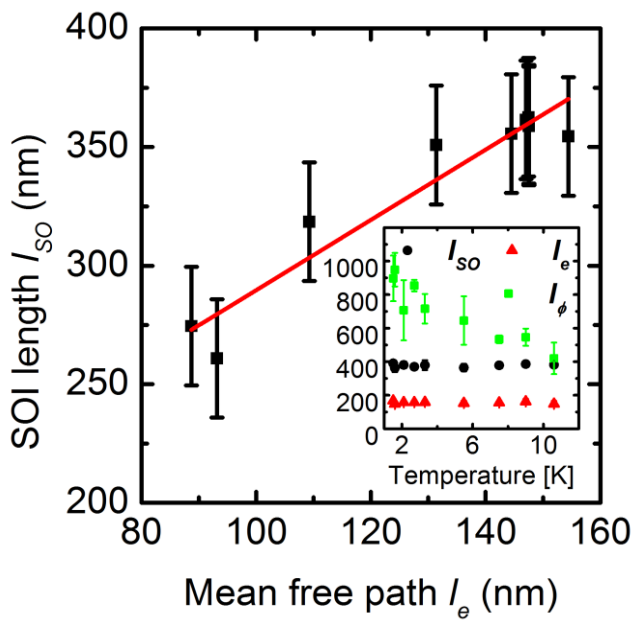

*Figure 6: $l_{SO}$-$l_e$ graph for the InAs wafer (B) at 1.7 K. Only a back gate was used to vary the transport parameters. Inset: The temperature dependence of the three characteristic lengths obtained from the same wafer matches the InAs/GaSb wafer (A) in the higher temperature range. The phase coherence length for wafer B is linear in temperature from 1.7 up to approximately 15 K, where it falls below the*

*SOI length (which is, as well as $l_e$, almost temperature independent) and consequently no WAL is observable.*

[1] M. Drndic, M. P. Grimshaw, L. J. Cooper, D. A. Ritchie, and N. K. Patel, Applied Physics Letters **70**, 481 (1997).
[2] R. J. Wagner, B. V. Shanabrook, M. J. Yang, and J. R. Waterman, Superlattices and Microstructures **21**, 95 (1997).
[3] M. J. Yang, C. H. Yang, B. R. Bennett, and B. V. Shanabrook, Physical Review Letters **78**, 4613 (1997).
[4] T. P. Marlow, L. J. Cooper, D. D. Arnone, N. K. Patel, D. M. Whittaker, E. H. Linfield, D. A. Ritchie, and M. Pepper, Physical Review Letters **82**, 2362 (1999).
[5] Y. Vasilyev, S. Suchalkin, K. von Klitzing, B. Meltser, S. Ivanov, and P. Kop'ev, Physical Review B **60**, 10636 (1999).
[6] K. Suzuki, S. Miyashita, and Y. Hirayama, Physical Review B **67**, 195319 (2003).
[7] C. Petchsingh, R. J. Nicholas, K. Takashina, N. J. Mason, and J. Zeman, Physical Review B **70**, 155306 (2004).
[8] K. Nilsson, A. Zakharova, I. Lapushkin, S. T. Yen, and K. A. Chao, Physical Review B **74**, 075308 (2006).
[9] C. Liu, T. L. Hughes, X. L. Qi, K. Wang, and S. C. Zhang, Phys Rev Lett **100**, 236601 (2008).
[10] I. Knez, R. R. Du, and G. Sullivan, Physical Review B **81**, 201301 (2010).
[11] I. Knez, R. R. Du, and G. Sullivan, Phys Rev Lett **107**, 136603 (2011).
[12] I. Knez, R. R. Du, and G. Sullivan, Phys Rev Lett **109**, 186603 (2012).
[13] I. Knez, C. T. Rettner, S. H. Yang, S. S. Parkin, L. Du, R. R. Du, and G. Sullivan, Phys Rev Lett **112**, 026602 (2014).
[14] C. Charpentier, S. Fält, C. Reichl, F. Nichele, A. Nath Pal, P. Pietsch, T. Ihn, K. Ensslin, and W. Wegscheider, Applied Physics Letters **103**, 112102 (2013).
[15] F. Nichele, A. N. Pal, P. Pietsch, T. Ihn, K. Ensslin, C. Charpentier, and W. Wegscheider, Phys Rev Lett **112**, 036802 (2014).
[16] E. M. Spanton, K. C. Nowack, L. Du, G. Sullivan, R. R. Du, and K. A. Moler, Phys Rev Lett **113**, 026804 (2014).
[17] L. Du, I. Knez, G. Sullivan, and R. R. Du, Phys Rev Lett **114**, 096802 (2015).
[18] V. S. Pribiag, A. J. Beukman, F. Qu, M. C. Cassidy, C. Charpentier, W. Wegscheider, and L. P. Kouwenhoven, Nat Nanotechnol **10**, 593 (2015).
[19] T. Li, P. Wang, H. Fu, L. Du, K. A. Schreiber, X. Mu, X. Liu, G. Sullivan, G. A. Csathy, X. Lin, and R. R. Du, Phys Rev Lett **115**, 136804 (2015).
[20] M. Karalic, S. Mueller, C. Mittag, K. Pakrouski, Q. Wu, A. A. Soluyanov, M. Troyer, T. Tschirky, W. Wegscheider, K. Ensslin, and T. Ihn, Physical Review B **94**, 241402 (2016).
[21] K. H. Kim, D. S. Um, H. Lee, S. Lim, J. Chang, H. C. Koo, M. W. Oh, H. Ko, and H. J. Kim, ACS Nano **7**, 9106 (2013).
[22] J. Luo, H. Munekata, F. F. Fang, and P. J. Stiles, Physical Review B **41**, 7685 (1990).
[23] G. L. Chen, J. Han, T. T. Huang, S. Datta, and D. B. Janes, Physical Review B **47**, 4084 (1993).
[24] J. P. Heida, B. J. van Wees, J. J. Kuipers, T. M. Klapwijk, and G. Borghs, Physical Review B **57**, 11911 (1998).
[25] T. Koga, J. Nitta, T. Akazaki, and H. Takayanagi, Phys Rev Lett **89**, 046801 (2002).
[26] D. Grundler, Phys Rev Lett **84**, 6074 (2000).
[27] C. Schierholz, T. Matsuyama, U. Merkt, and G. Meier, Physical Review B **70**, 233311 (2004).
[28] K. H. Kim, H. C. Koo, J. Chang, Y. S. Yang, and H. J. Kim, Ieee Transactions on Magnetics **50**, 18 (2014).
[29] C. Brüne, A. Roth, H. Buhmann, E. M. Hankiewicz, L. W. Molenkamp, J. Maciejko, X.-L. Qi, and S.-C. Zhang, Nature Physics **8**, 486 (2012).
[30] P. D. Dresselhaus, C. M. Papavassiliou, R. G. Wheeler, and R. N. Sacks, Phys Rev Lett **68**, 106 (1992).
[31] R. J. Elliott, Physical Review **96**, 266 (1954).
[32] F. Nichele, M. Kjaergaard, H. J. Suominen, R. Skolasinski, M. Wimmer, B.-M. Nguyen, A. A. Kiselev, W. Yi, M. Sokolich, M. J. Manfra, F. Qu, A. J. A. Beukman, L. P. Kouwenhoven, and C. M. Marcus, Physical Review Letters **118**, 016801 (2017).
[33] S. Hikami, A. I. Larkin, and Y. Nagaoka, Progress of Theoretical Physics **63**, 707 (1980).
[34] T. Matsuyama, R. Kuersten, C. Meißner, and U. Merkt, Physical Review B **61**, 15588 (2000).
[35] A. C. H. Rowe, J. Nehls, R. A. Stradling, and R. S. Ferguson, Physical Review B **63**, 201307 (2001).
[36] C. Morrison, J. Foronda, P. Wiśniewski, S. D. Rhead, D. R. Leadley, and M. Myronov, Thin Solid Films **602**, 84 (2016).
[37] G. Bergmann, Physics Reports **107**, 1 (1984).
[38] G. Dumpich and A. Carl, Physical Review B **43**, 12074 (1991).
[39] Y. Ho Park, H.-j. Kim, J. Chang, S. Hee Han, J. Eom, H.-J. Choi, and H. Cheol Koo, Applied Physics Letters **103**, 252407 (2013).
[40] K.-H. Kim, H.-j. Kim, H. C. Koo, J. Chang, and S.-H. Han, Applied Physics Letters **97**, 012504 (2010).
[41] S. Birner, T. Zibold, T. Andlauer, T. Kubis, M. Sabathil, A. Trellakis, and P. Vogl, Ieee Transactions on Electron Devices **54**, 2137 (2007).
[42] X. J. Hao, T. Tu, G. Cao, C. Zhou, H. O. Li, G. C. Guo, W. Y. Fung, Z. Ji, G. P. Guo, and W. Lu, Nano Lett **10**, 2956 (2010).
[43] I. Žutić, J. Fabian, and S. Das Sarma, Reviews of Modern Physics **76**, 323 (2004).